\documentclass[]{aa}  
\usepackage[dvips]{graphicx}

\usepackage[usenames, dvipsnames]{xcolor}
\usepackage{color}
\usepackage{txfonts}
\usepackage{natbib}
\usepackage{psfrag}
\usepackage{bm}

\bibpunct{(}{)}{;}{a}{}{,} 

\def\be{\begin{equation}}
\def\ee{\end{equation}}
\def\kms{{\rm\,km\,s^{-1}}}
\def\kmskpc{{\rm\,km\,s^{-1}\,{kpc}^{-1}}}
\def\pc{{\rm\,pc}}

\def\Rsun{{\rm\,R_\odot}}

\def\deg{{^\circ}}

\def\kpc{{\rm\,kpc}}

\def\1s{{1$\sigma$}}
\def\2s{{2$\sigma$}}
\def\3s{{3$\sigma$}}

\def\dex{{\rm\,dex}}

\newcommand{\bea}	{\begin{array}}
\newcommand{\eea}	{\end{array}}
\newcommand{\ben}	{\begin{eqnarray}}
\newcommand{\een}	{\end{eqnarray}}
\newcommand{\bsq}	{\begin{mathletters}}
\newcommand{\esq}	{\end{mathletters}}

\newcommand{\Ro}	{R_0}

\newcommand{\vo}	{v_0}

\newcommand{\vphi}	{v_{\phi}}
\newcommand{\vr}	{v_{R}}

\newcommand{\Ws} 	{W_\odot}
\newcommand{\Vs} 	{V_\odot}
\newcommand{\Us} 	{U_\odot}

\newcommand{\mh}{[\rm M/\rm H]}

\newcommand{\feh}{[\rm Fe/\rm H]}

\def\aap{A\&A}

\def\2s{2-$\sigma$}
\def\3s{3-$\sigma$}

\begin{document}
 \title{Asymmetric metallicity patterns \\ in the stellar velocity space with RAVE}
\titlerunning{Metallicity patterns in the stellar velocity space}

   \author{T.~Antoja
          \inst{1,2}\fnmsep\thanks{ESA Research Fellow.}
          \and
           G.~Kordopatis\inst{3,4}
          \and
          A.~Helmi\inst{5} 
          \and 
          G.~Monari\inst{6,7} 
          \and
          B.~Famaey\inst{6}
           \and
          R.~F.~G.~Wyse\inst{8}
           \and
          E.~K.~Grebel\inst{9}
           \and	
          M.~Steinmetz\inst{3}
          \and
          J.~Bland-Hawthorn\inst{10}
          \and
          B.~K.~Gibson\inst{11}          
          \and
          O.~Bienaym\'e\inst{6}
          \and 
          J.~F.~Navarro\inst{12}
          \and
          Q.~A.~Parker\inst{13}
          \and
          W.~Reid\inst{14,15}
          \and
          G.~Seabroke\inst{16}
          \and 
          A.~Siebert\inst{6}
         \and
       {   A.~Siviero\inst{17}}
         \and
          T.~Zwitter\inst{18}
          }

   \institute{Directorate of Science, European Space Agency (ESA-ESTEC), PO Box 299, 2200 AG Noordwijk, The Netherlands
                      \and
        {Dept. FQA, Institut de Ciencies del Cosmos (ICCUB), Universitat de Barcelona (IEEC-UB), Marti Franques 1, E08028 Barcelona, Spain\\}
          \email{tantoja@fqa.ub.edu}
         \and
         Leibniz-Institut f\"ur  Astrophysik Potsdam (AIP), An der Sternwarte 16, 14482 Potsdam, Germany
 \and
{Universit\'e C\^{o}te d'Azur, Observatoire de la C\^{o}te d'Azur, CNRS, Laboratoire Lagrange, Bd de l'Observatoire, CS
34229, 06304
Nice Cedex 4, France}
\and
 Kapteyn Astronomical Institute, University of Groningen, PO Box 800, NL-9700 AV Groningen, the Netherlands
 \and
 Observatoire astronomique de Strasbourg, Universit\'e de Strasbourg, CNRS UMR 7550, 11 rue de l'Universit\'e, 67000 Strasbourg, France
   \and
     {The Oskar Klein Centre for Cosmoparticle Physics, Department of Physics, Stockholm University, AlbaNova, 10691 Stockholm, Sweden}
     \and 
  Department of Physics and Astronomy, Johns Hopkins University, 3400 N. Charles St, Baltimore, MD 21218, USA
   \and
  Astronomisches Rechen-Institut, Zentrum f\"ur Astronomie der Universit\"at Heidelberg, M\"onchhofstr.\ 12-14, 69120 Heidelberg, Germany
  \and
 Sydney Institute for Astronomy, School of Physics A28, University of Sydney, NSW 2006, Australia
 \and
  E.A. Milne Centre for Astrophysics, University of Hull, Hull, HU6 7RX, United Kingdom
   \and
  Senior CIfAR Fellow. University of Victoria. Victoria, BC Canada V8P 5C2
  \and
  Department of Physics, Chong Yuet Ming Physics Building, The University of Hong Kong, Hong Kong
  \and
  Department of Physics and Astronomy, Macquarie University, Sydney, NSW 2109, Australia
  \and
  Western Sydney University, Locked Bag 1797, Penrith South DC, NSW 1797, Australia
  \and
  Mullard Space Science Laboratory, University College London, Holmbury St Mary, Dorking, RH5 6NT, United Kingdom
  \and
    {Dipartimento di Fisica e Astronomia Galileo Galilei, Universita’ di Padova, Vicolo dell'Osservatorio 3, I-35122 Padova, Italy}
\and
Faculty of Mathematics and Physics, University of Ljubljana, 1000 Ljubljana, Slovenia
             }

%
%
%
%

   \date{Received XX; accepted XX}


  \abstract{The chemical abundances of stars encode information on their place and time of origin. Stars formed together in e.g. a cluster, should present chemical homogeneity. Also disk stars influenced by the effects of the bar and the spiral arms might have distinct chemical signatures depending on the type of orbit that they follow, e.g. from the inner versus outer regions of the Milky Way. }
{We explore the correlations between velocity and metallicity and the possible distinct chemical signatures of the velocity over-densities of the local Galactic neighbourhood.}
{We use the large spectroscopic survey RAVE and the Geneva Copenhagen Survey. We compare the metallicity distribution of regions in the velocity plane  ($\vr,\vphi$) with that of their symmetric counterparts ($-\vr,\vphi$). We expect similar metallicity distributions if there are no tracers of a sub-population (e.g., a dispersed cluster, accreted stars), if the disk of the Galaxy is axisymmetric, and if the orbital effects of the bar and the spiral arms are weak.}
{We find that the metallicity-velocity space of the solar neighbourhood is highly patterned. 
A large fraction of the velocity plane shows differences in the metallicity distribution when comparing symmetric $\vr$ regions. 
{The typical differences in the median metallicity are of $0.05\dex$ with  statistical significance of at least $95\%$, and with values up to $0.6\dex$}. 
For stars with low azimuthal velocity $\vphi$, the ones moving outwards in the Galaxy have on average higher metallicity than those moving inwards. These include stars in the Hercules and Hyades moving groups and other velocity branch-like structures. For higher $\vphi$, the stars moving inwards have higher metallicity than those moving outwards. 
{We have also discovered a positive gradient in $\vphi$ with respect to metallicity at high metallicities, apart from the two known positive and negative gradients for the thick and thin disks, respectively.}}
{The most likely interpretation of  {the metallicity asymmetry} is that it is due to the {orbital effects of the} Galactic bar {and the  radial metallicity gradient of the disk.} We present a simulation that supports this idea. }

   \maketitle

\section{Introduction}\label{intro}

Galaxies evolve through a complex intermix of internal and
external mechanisms. For the disk of our Galaxy, we can identify
several processes that give shape to its current structure,
kinematics, {and} to its chemical and population properties. All these
processes are expected to leave kinematic imprints and a fossil record
in its velocity distribution \citep{Freeman2002}. 
{For example, the presence of a bar (a non-axisymmetric gravitational component) leaves an imprint in the velocity distribution, making it asymmetric with respect to the Galactic cylindrical coordinate $\vr$ \citep{Dehnen2000}. In general, asymmetries in $\vr$ are a signature of breakdown of axisymmetry. This could arise either from incomplete phase mixing or from the presence of non-axisymmetric components of the potential.}

{Processes that play a role in sculpting the velocity distribution are the following.} Firstly, star formation takes place in gaseous complexes,
forming stellar clusters that sooner or later are disrupted due to the
tidal stripping and internal evolution
{{(e.g. \citealt{Spitzer1987,Baumgardt2003,Bible2008})}. When these clusters dissolve
they can leave imprints in the velocity space while being
already quite dispersed in space \citep{Skuljan1997}. Secondly, the {orbits of} the stars in the disk {are} perturbed by the spiral arms and the bar, {especially} through mechanisms such as {driven eccentricity and trapping at the resonances}
{\citep{Dehnen2000,Kalnajs1991,Sridhar1996}}, or radial mixing, making stars migrate
 from the radius where they were born
(e.g. \citealt{Sellwood2002,Roskar2008}, {\citealt{Schonrich2009}}, \citealt{Minchev2010b}). 
{This} resonant interaction can {make} stars appear to be concentrated in over-densities (dynamical streams) in the velocity plane 
(e.g., {\citealt{Dehnen2000}}, \citealt {Antoja2009,Quillen2005}).
Satellite galaxies also perturb the orbits in the  disk
{\citep[e.g.,][]{Quinn1993,Kazantzidis2008,Purcell2011}}, 
which can also
induce kinematic substructure
\citep{Quillen2009}. Furthermore, perturbing satellites can  leave
stars behind in the disk 
that will define velocity clumps \citep{Villalobos2009}. 


An important issue is how to distinguish between the different processes. This is necessary to ultimately decode the relative importance and roles of these mechanisms in the formation and evolution of our Galaxy. In addition to the phase-space stellar distribution of the disk, stellar ages and chemical information are key elements in this decoding. While the ages of stars are difficult to measure \citep[e.g.][]{Soderblom2010,Kordopatis2016}, we can use the measured chemical abundances as indicators of the time of formation of the stars and their place of origin.

Several studies have used Str\"omgren and high-resolution spectroscopy to find clues about the origin of some velocity over-densities. For instance, the chemical homogeneity of the stars in the group HR1614 found in \citet{Feltzing2000} and \citet{DeSilva2007} indicates that it is a remnant of a dispersed cluster. Other moving groups, e.g., Hyades, Sirius, Pleiades and Hercules, were {at one time} related to dispersed clusters. However, the large dispersions in age and metallicity {of their stars} \citep{Famaey2005,Helmi2006,Antoja2008, Ramya2016} discarded this hypothesis. 
This  favours an origin related to the resonances of the bar and the spiral arms. 

In the case of resonant effects, {the stellar chemistry can give clues} on the type of orbit that the stars follow. For example, if there is/was a metallicity gradient in the Galaxy, as measured, e.g., in \citet{Gazzano2013, Boeche2013,Genovali2014,Hayden2014},
or modelled in \citet{Minchev2013}, 
then orbits from the inner versus outer regions of the Milky Way should { have different metallicities, therefore, leaving} different signatures {in the combined space of chemistry and phase-space}. These would potentially help us to understand which resonance causes each of the over-densities, {breaking the current} degeneracies (several groups can be explained by either effects of the bar or the spiral arms, \citealt{Antoja2009,Antoja2011}). 

\citet{Bovy2010} have explored this  by comparing the metallicity distribution of the main known moving groups to {that} of the background population. They found that, in general, there are no distinguishable chemical patterns in the main over-densities, which argues in favour of an origin related to transient perturbations more than to the long lasting effect of resonances. 
But they also pointed out that stellar migration could have erased some of the expected metallicity signatures.

 Some moving groups in the local neighbourhood seem to have an extra-galactic origin \citep{Helmi1999,WyliedeBoer2010}. Other groups remain controversial. For instance, based on chemical abundances, the Arcturus moving group was thought to be a remnant of an ancient merger event \citep{Navarro2004} but other studies showed that its chemical and age distribution point to an internal origin \citep{Williams2009,Bensby2014}.


Here we take advantage of the large spectroscopic survey RAVE \citep{Steinmetz2006} to explore the correlations between velocity and metallicity and the possible distinct chemical signatures of the kinematic over-densities of the local neighbourhood. 
We use a novel approach: we test the hypothesis that the metallicity distribution of a certain region of velocity space, e.g., in Galactic cylindrical coordinate velocities ($\vr,\vphi$), is compatible with the one for its symmetric region ($-\vr,\vphi$). This should be the case {under the hypothesis of axisymmetry.} {But, again, this can be broken by i) {incomplete phase-mixing}  (e.g., a sub-population that is not yet phase-mixed such as a disrupted cluster or an accreted satellite),} {ii) the non-axisymmetries of the potential (e.g., bar and spiral arms).} {In the first case, the sub-population would form a region in the velocity space with a distinct metallicity. In the second one, stars following orbits of opposite signs of $\vr$ would not have  the same guiding radii, and thus, have different metallicities given by the disk metallicity gradient.} This methodology has the advantage that it does not use a comparison with the background stars, which might lead to inaccurate results if the field is dominated by asymmetries and groups. We find for the first time a significant asymmetry in the metallicities for positive and negative $\vr$. 

We describe the samples in Section~\ref{sample}. In Section~\ref{met} we study the metallicity {distribution } of the velocity plane and the known moving groups. In Section~\ref{assym} we compare the metallicity distributions of all regions in velocity space with their symmetric counterparts {in $\vr$}. Finally, in Sections~\ref{assymvphi} and~\ref{gradients} we study the global metallicity asymmetries and gradients seen in the velocity plane. We discuss our findings and conclude in Section~\ref{concl}.


\section{Observational samples}\label{sample}



{We use the RAVE Data Release 5 (DR5, \citealt{Kunder2016}) which includes the new metallicity calibration of \citet{Kordopatis2015}.}
The stellar parallaxes were obtained through the Bayesian distance-finding method of \citet{Binney2014}.
First we select stars with i) SNR better than 20, ii) the first morphological flag indicating that they are normal stars \citep{Matijevic2012}, and iii) converged algorithm of computation of the physical parameters \citep[``algo\_conv'' flag = 0, see][]{Kordopatis2013}. From these, we further select those in a cylinder with radius of $0.5\kpc$ and total height of $1\kpc$ centred on the Sun's position. This results in a sample of {166015} stars with
6D phase-space information and $\mh$.
We use proper motions from UCAC4 \citep{Zacharias2013} but we have tested that our results do not change significantly if using PPMXL \citep{Roeser2010}. The median relative distance error is {$31\%$} and the median error in $\vr$ and $\vphi$ are around 7 and $5\kms$, respectively.
The median error in metallicity $\mh$ is $0.1\dex$.

We also use the Geneva-Copenhagen survey (hereafter GCS, \citealt{Holmberg2009}) and the new metallicity re-analysis by \citet{Casagrande2011}. We select stars with the flag of good quality index and with available 3D velocities and $\mh$. This sample has 11379 stars.
The median relative error in distance for this sample is $6\%$ and the median {errors} in $\vr$ and $\vphi$ are 1.5 and $1.2\kms$, respectively. This is a much more local sample, with a median distance from the Sun of $71\pc$. Note that the metallicity scales of RAVE and GCS are not necessarily the same, and some small offset is expected between the measured metallicities of the two surveys \citep[see][their Fig.~12, for a comparison of the GCS to the RAVE metallicities]{Kordopatis2013}.


Following \citet{Reid2014}, we assume that the Sun is at $X =-8.34\kpc$ and the circular velocity at the Sun of is $\vo=240\kms$. For the velocity of the Sun with respect to the Local Standard of Rest we adopt $(\Us, \Vs, \Ws)=(10,12,7)\kms$ \citep{Schonrich2010}. The resulting value of $(\vo+\Vs)/\Ro$ is $30.2\kmskpc$, which is compatible with that from the reflex motion of the Sgr A* of $30.2 \pm 0.2 \kmskpc$ \citep{Reid2004}. With these values, 
we compute the cylindrical velocities of the stars in our samples: $\vr$ (positive towards the Galactic centre, in consonance with the usual $U$ velocity component) and $\vphi$ (towards the direction of rotation).


\section{Metallicity patterns in the local  velocity plane}\label{met}

    \begin{figure*}
   \centering
   \includegraphics[height=0.3\textheight]{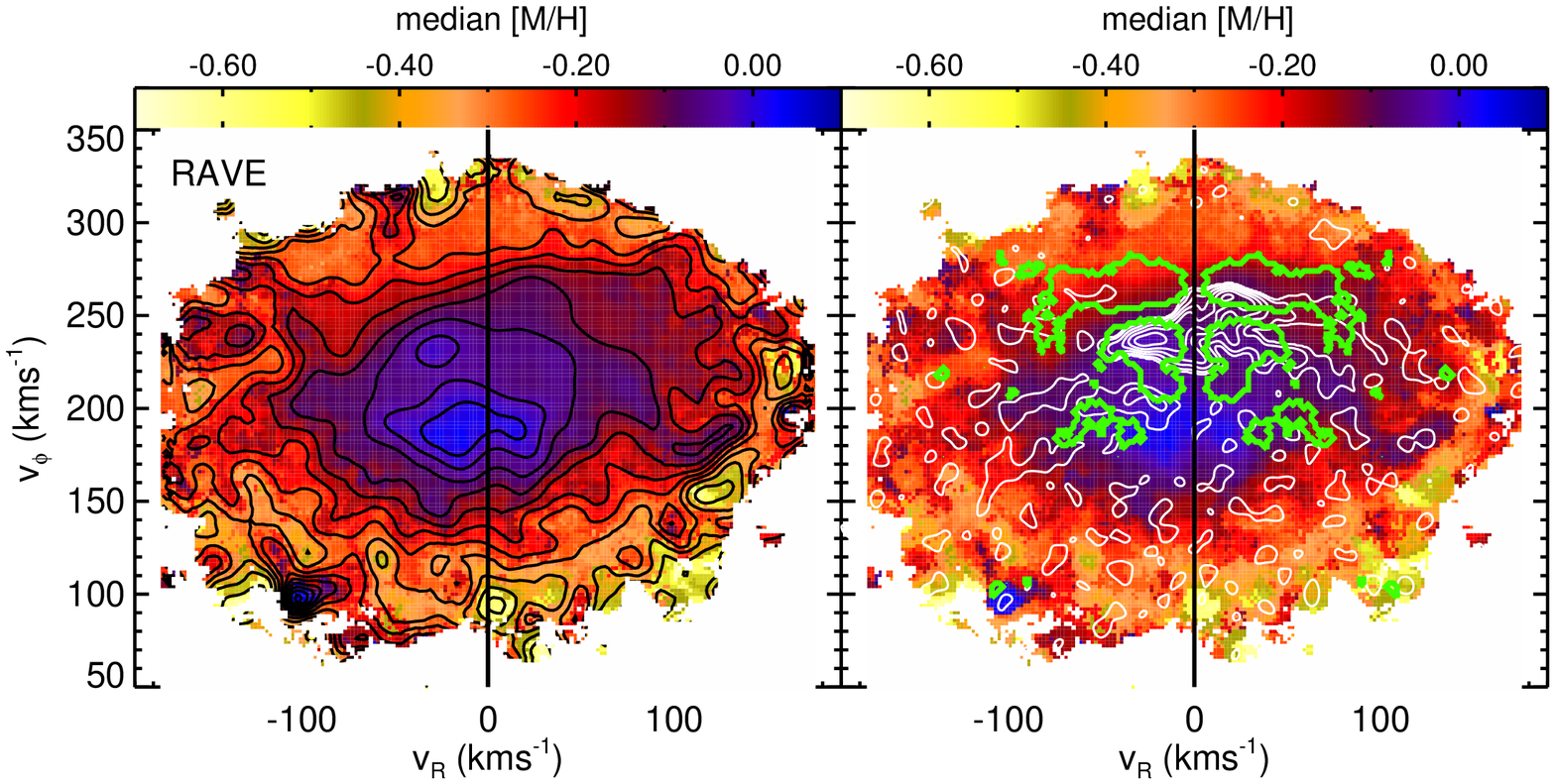}\hspace{-1.6cm}
 \includegraphics[height=0.3\textheight]{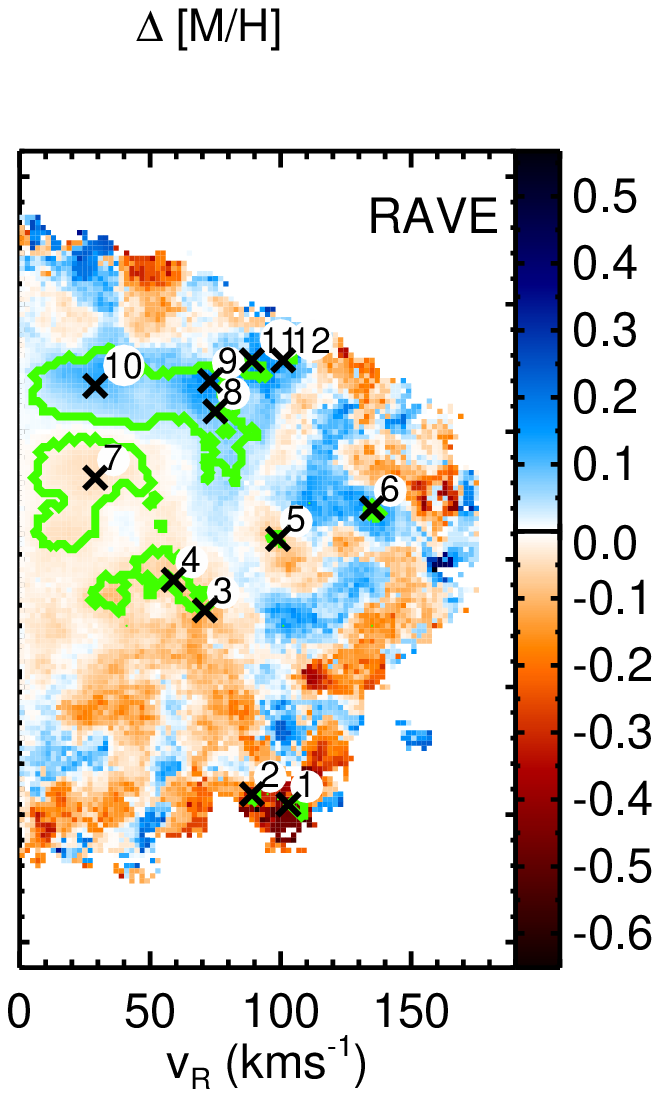}

  \includegraphics[height=0.3\textheight]{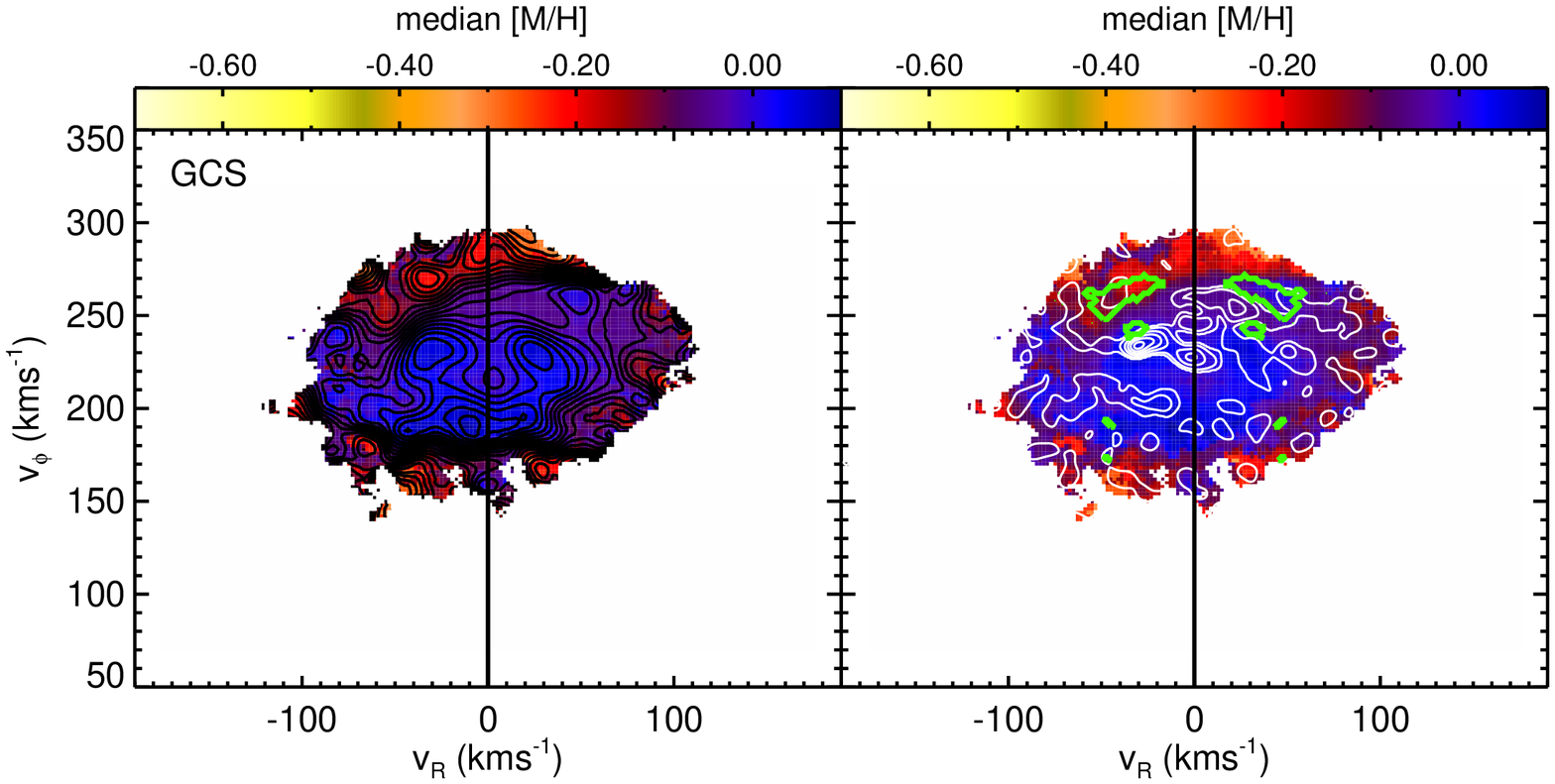}  \hspace{-1.6cm} 
  \includegraphics[height=0.3\textheight]{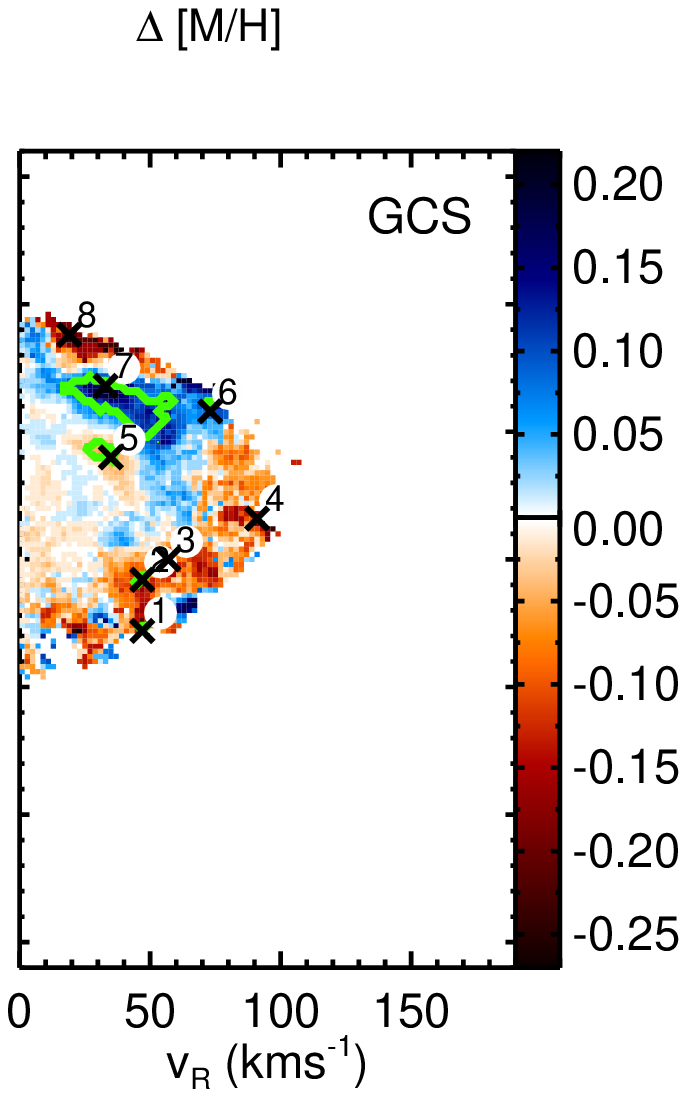}
   
  \caption{{\em Top:} Median metallicity in the velocity plane of the local neighbourhood of the RAVE sample. The {\it left} {panel shows 25 logarithmically spaced contours in the range spanned by the metallicity, i.e. [-0.70,0.06].} {The {\it middle} panel is the same as the left one but } shows contours (white) of the wavelet transform that mark the main over-densities in the velocity plane, and contours (green) that indicate {significant metallicity differences} ($95\%$ confidence) from its symmetric $\vr$ region. The {\it right} panel shows the differences in the median metallicity for positive and negative $\vr$. The green contours {are the same as for the middle panel.} 
The numbered labels are points identified as depicting clear differences in metallicity, and whose characteristics are given in Table~\ref{tRAVE}. {\it Bottom:} Same for the GCS sample, with the labelled points given in Table~\ref{tGCS}. {The {\it left} panel shows now 25 logarithmically spaced contours in the range of [-0.34,0.1]. }}
         \label{metvrvphi}
   \end{figure*} 

Here we study the stellar metallicity $\mh$ as {a} function of position in the velocity plane. We start from a grid of points in the velocity space $\vr-\vphi$ separated by $2\kms$. We assign to each point the median metallicity $\mh$ of the stars in a circle of $10\kms$. We see no differences in the results when we compute the mean metallicity instead of the median. We only consider points of the grid with at least 10 stars. We compute the error on the median metallicity through bootstrapping of 10000 samples for each point. Since the distribution of the bootstrapped median $\mh$ is not necessarily symmetric, we work with confidence limits, in particular, the $95\%$ confidence range.

The top left panel of Fig.~\ref{metvrvphi} shows the median metallicity in the velocity plane of the RAVE sample. The metallicity distribution is not uniform. The more metal rich regions are concentrated in the centre of the distribution (blue colours) while the metal poor ones are distributed in the outer parts (yellow and orange). This is likely the effect of the young stars having less velocity dispersion. We also see a highly patterned metallicity distribution: there are structures of different sizes and shapes (e.g., rounded or with branch shape) that present higher/lower metallicity than its surroundings.

{To examine if there is any correlation between the metallicity patterns and the known kinematic over-densities, in the middle panel, we superpose in white the contours of the wavelet transform (WT, \citealt{Starck2002}). 
The wavelet transform is a decomposition of a signal into scale-related views and thus shows over-densities of certain scales.
}
{Here we show
the scale of $6\kms$ which has been demonstrated to highlight the main known velocity over-densities (see Antoja et al. 2008, 2012, 2015b which also describe the method). The main features are: 
\begin{itemize}
\item an elongated feature of high metallicitiy (blue colours) that coincides roughly with the Hercules stream at $(\vr, \vphi)\sim(-50,200)\kms$
\item the region of the Hyades stream $\sim(-20,240)\kms$
\item an elongated structure, hereafter called branch 1, ranging from  $\sim{(50,220)}\kms$ to $\sim(130,190)\kms$ (purple color) that includes the Wolf 630 and Dehnen 98 moving groups (Antoja et al. 2012) and also referred to as "the horn"
\item a velocity feature at $\sim(70,250)\kms$ (hereafter branch 2), which includes part of the Sirius group, with a metallicity higher than its surroundings.
\end{itemize}
}
  

The bottom panels of Fig.~\ref{metvrvphi} are for the GCS sample, where we observe a similar metallicity pattern, with also the streams of Hercules, Hyades and {branch 1} being the {most} metal rich. {Note that the colour scale used for these plots is the same but the GCS sample shows higher metallicities. This is because this sample is  biased towards velocities close to the LSR and the highest metallicity part of the solar neighbourhood. }

\section{Metallicity asymmetries in the velocity plane}\label{assym}

 {The chemical properties of the solar neighbourhood should be symmetrical in $\vr$ if there is axisymmetry. But we see in Fig.~\ref{metvrvphi} that the metallicity of the velocity plane is clearly asymmetric with respect to $\vr=0$. Here} we test the hypothesis that the metallicity distribution, and specifically its median, is the same for a point $(v_{R1},v_{\phi1})$ and its symmetric point  $(v_{R2},v_{\phi2})=(-v_{R1},v_{\phi1})$. 
To {do it} we compute the difference between the median metallicity {of these two points}: 
\be
\Delta  \mh= \mh_1 - \mh_2
\ee
{where we use the subscript 1 for the point at positive $\vr$ and 2 for the symmetric one at negative $\vr$.} This quantity is shown in the right panels of Fig.~\ref{metvrvphi}. The blue colours show regions where the right part of the velocity distribution ($\vr>0$) is more metal rich compared to the left side, and red colours {show} the opposite.  
{Green lines show the velocity regions for which the statistical difference is greater than 95\%. These lines are are reproduced in the middle and right panels of Fig.~\ref{metvrvphi}. }

   \begin{table*}
  \setlength{\tabcolsep}{1.1pt}
\caption{Metallicity differences in the symmetric velocity points for the RAVE sample marked in the right top panel of Fig.~\ref{metvrvphi}. Columns show: 1) structure (pixel) ID number, 2)  absolute value of the Galactic radial velocity of the pixels at positive and negative $\vr$, 3) azimuthal velocity of both pixels, 4) median metallicity of the pixel at {positive} $\vr$, 5) $95\%$ confidence range for the median metallicity of the pixel at {positive} $\vr$, 6) {same as 4 but for} the pixel at {negative} $\vr$, 7) {same as 5 but for} the pixel at {negative} $\vr$, 8) difference between median metallicities: $\Delta  \mh= \mh_1 - \mh_2$, 9) approximate significance of the difference (Eq.~\ref{eq2}), 10) number of stars in the circle of $10\kms$ around the pixel at {positive} $\vr$, 11) {same as 10 but for} the pixel at {negative} $\vr$, 12) associated moving group.}\label{tRAVE}      
 \centering          
 \begin{tabular}{cccccccccrrl}     
 \hline\hline    
 ID& $|v_{R}|$   &$\vphi$&  $\mh_1$ & $95\%$ range$_1$&  $\mh_2$ & $95\%$ range$_2$& $\mh_1-\mh_2$ &$\sigma$  &  $N_1$ &  $N_2$&moving group\\
 &($\kms$)    &($\kms$) &  (dex)& (dex)&   (dex)   &(dex) &(dex) & &   & & \\\hline 

 1&  103&  104&  -0.62 &[  -0.86 ,  -0.20]&   0.04&[  -0.18 ,  0.39]& -0.65&   3.5&    16&    12&\\
 2&   89&  108&  -0.52 &[  -0.95 ,  -0.30]&  -0.06&[  -0.25 ,  0.06]& -0.46&   2.4&    17&    20&\\
 3&   71&  180&  -0.18 &[  -0.26 ,  -0.13]&  -0.08&[  -0.12 , -0.05]& -0.10&   2.5&   166&   416&Hercules\\
 4&   59&  192&  -0.10 &[  -0.12 ,  -0.08]&  -0.03&[  -0.06 , -0.00]& -0.07&   3.7&   419&   822&Hercules\\
 5&   99&  208&  -0.19 &[  -0.24 ,  -0.14]&  -0.10&[  -0.13 , -0.07]& -0.09&   3.4&   229&   251&$\sim$Hercules ($\vr<0$), branch 1($>0$) \\
 6&  135&  220&  -0.11 &[  -0.18 ,  -0.02]&  -0.27&[  -0.31 , -0.21]&  0.16&   3.1&    70&    42&extension branch 2 \\
 7&   29&  232&  -0.06 &[  -0.07 ,  -0.05]&  -0.01&[  -0.02 , -0.01]& -0.04&   7.9&  6428&  7469&Hyades ($\vr<0$), branch 1 ($>0$)\\
 8&   75&  258&  -0.11 &[  -0.13 ,  -0.09]&  -0.20&[  -0.22 , -0.16]&  0.08&   4.2&   848&   416&branch 2\\
 9&   73&  270&  -0.11 &[  -0.14 ,  -0.08]&  -0.27&[  -0.30 , -0.23]&  0.17&   7.2&   356&   225&$\sim$branch 2\\
10&   29&  268&  -0.09 &[  -0.10 ,  -0.09]&  -0.21&[  -0.23 , -0.19]&  0.12&  10.3&  3227&  1509&$\sim$Sirius, branch 2\\
11&   89&  278&  -0.15 &[  -0.23 ,  -0.10]&  -0.33&[  -0.41 , -0.27]&  0.19&   3.4&    59&    64&\\
12&  101&  278&  -0.16 &[  -0.26 ,  -0.10]&  -0.30&[  -0.41 , -0.27]&  0.14&   2.1&    39&    55&\\

\hline
 \end{tabular}
   \end{table*}

   \begin{table*}
  \setlength{\tabcolsep}{1.1pt}
\caption{Same as Table~\ref{tRAVE} but for the GCS sample.}\label{tGCS}      
 \centering          
 \begin{tabular}{cccccccccrrl}     
 \hline\hline    
ID  & $|v_{R}|$    &$\vphi$&  $\mh_1$ & $95\%$ range$_1$&  $\mh_2$ & $95\%$ range$_2$& $\mh_1-\mh_2$ &$\sigma$  &  $N_1$ &  $N_2$&moving group\\
 &($\kms$)   &($\kms$) &  (dex)& (dex)&   (dex)   &(dex) &(dex) & &   & & \\\hline 
GCS-1&   47&    172&  -0.23 &[  -0.43 ,  -0.16]&  -0.07&[  -0.14 ,  0.08]& -0.16&   1.6&    11&    18 &Hercules \\
GCS-2&   47&    192&  -0.09 &[  -0.17 ,  -0.03]&   0.05&[   0.00 ,  0.08]& -0.14&   3.7&    25&    55&Hercules\\
GCS-3&   57&    200&  -0.13 &[  -0.22 ,  -0.05]&   0.02&[  -0.02 ,  0.05]& -0.15&   3.0&    31&    86&Hercules\\
GCS-4&   91&    216&  -0.14 &[  -0.32 ,  -0.05]&   0.06&[  -0.04 ,  0.09]& -0.20&   2.0&    18&    14&Hercules ($\vr<0$), branch 1 ($>0$) \\
GCS-5&   35&    240&  -0.02 &[  -0.04 ,   0.00]&   0.04&[   0.02 ,  0.05]& -0.06&   4.7&   435&   647& $\sim$ Hyades \\
GCS-6&   73&    258&  -0.05 &[  -0.08 ,   0.04]&  -0.16&[  -0.26 , -0.08]&  0.11&   2.2&    50&    26& branch 2 \\
GCS-7&   33&    268&  -0.01 &[  -0.09 ,   0.03]&  -0.23&[  -0.27 , -0.20]&  0.22&   5.4&   102&    60& $\sim$Sirius, branch 2 \\
GCS-8&   19&    288&  -0.29 &[  -0.33 ,  -0.16]&  -0.05&[  -0.10 ,  0.03]& -0.24&   4.1&    23&    15&\\
 \hline
 \end{tabular}
 \end{table*}

We have chosen some representative points within these regions. 
In Tables~\ref{tRAVE} and \ref{tGCS}, for the RAVE and GCS samples, respectively, we indicate for each pair of points the difference in metallicity {and its significance
computed as\footnote{Note that this gives only an approximate idea of the significance of the detected signal since the bootstrapped samples may not have a symmetric distribution of median metallicities.}:
\be
\label{eq2}
\sigma \equiv \frac{\Delta\mh}{\sqrt{\sigma_1^2+\sigma_2^2}},
\ee
where $\sigma_1$ and $\sigma_2$ are the standard deviation of the bootstrapped {medians} at positive and negative $\vr$, respectively.}
{For RAVE (Table~\ref{tRAVE})} the {typical} metallicity differences are around 0.1 dex. 
Considering all points of the grid where the differences are significant (inside the green contours), the values range from
{0.009 to 0.65} with a median of {$0.05\dex$}. For GCS, they range between 0.03 and 0.27, with a median of $0.12\dex$. Note that the error on the individual stellar metallicity of  RAVE  is around 0.1 dex \citep{Kordopatis2013}. However, due to the high number of stars, the median metallicity can be determined with higher precision. All cases in Table~\ref{tRAVE} have $\sigma>2$ since we only listed points with differences of at least the $95\%$ confidence level, and we also find points with metallicity discrepancies that are 3, 4 and up to 10$\sigma$ significant. 


{The main velocity regions with $\bf{\mh}$ asymmetry are:
\begin{itemize}
\item The stream of Hercules (structure at pixel IDs {3, 4 \& 5}) with higher metallicity than the symmetric counterpart at $\vr>0$, with differences in $\mh$ up to {$0.1\dex$}.
\item The Hyades stream (ID {7}) with slightly higher metallicity ({$0.04\dex$}) than its symmetric counterpart (approximately coinciding with branch 1).
\item The large region in the upper part of the velocity distribution (IDs 8, 9 \& 10) that is up to {$0.17\dex$} more metal rich for $\vr>0$ compared to $\vr<0$. This region coincides partially with the Sirius group and branch 2. 
\item Other smaller regions with higher differences in $\mh$ such as those indicated with IDs 1, 2, 11 and 12. These do not correspond to any well-known kinematic over-densities. 
\end{itemize}}


For the GCS (Fig.~\ref{metvrvphi}, bottom), there
are fewer regions of significant metallicity difference since there are much fewer stars in the sample. The
significant regions show the same trends {as for RAVE}. For instance, pixel IDs {GCS-3, GCS-4, GCS-6
and GCS-7} are equivalent {or very close } to those with RAVE {IDs 4, 5, 8 and 10}, respectively. 

    \begin{figure}
   \centering
 \includegraphics[width=0.38\textwidth]{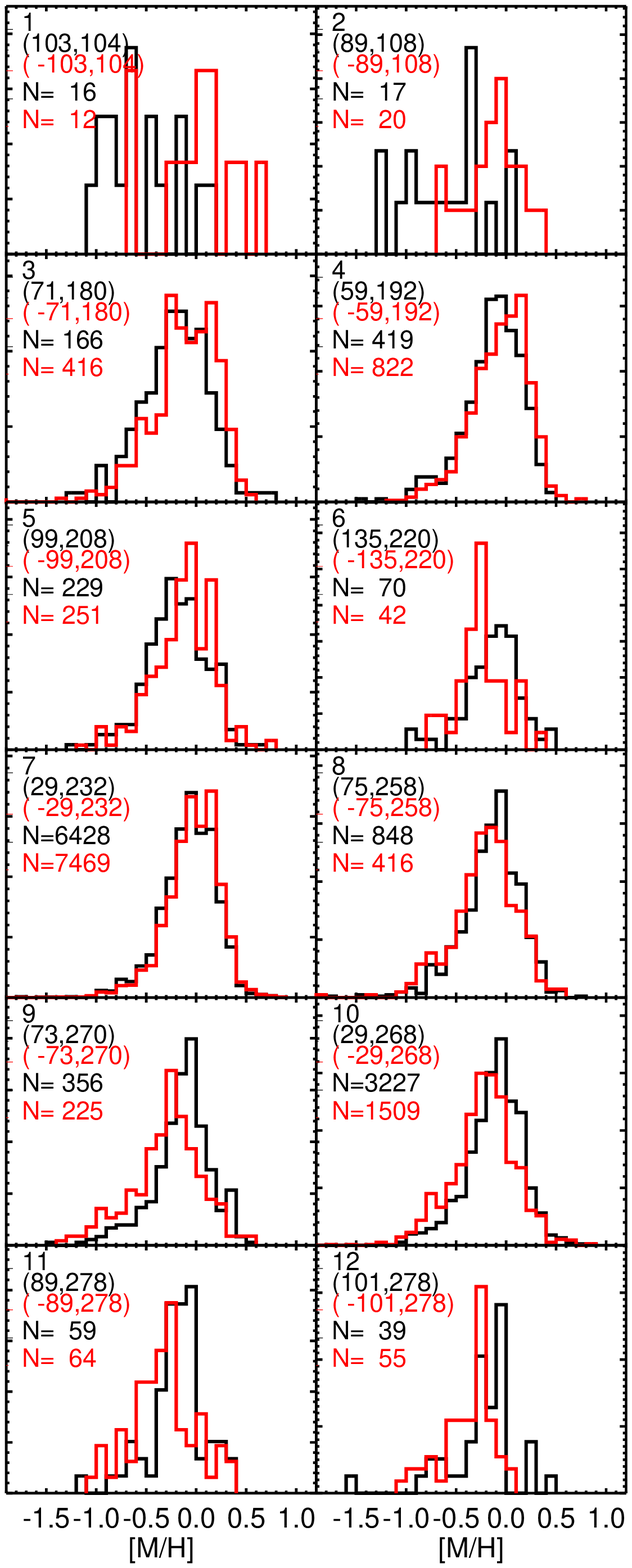} 
  \caption{Metallicity distributions of the locations identified in velocity space where the median metallicity is different with $95\%$ confidence between regions with opposite $\vr$ for the RAVE sample. The panels are numbered as in the right top panel of Fig.~\ref{metvrvphi}. Black and red histograms are normalized and correspond to the distributions for positive and negative $\vr$, respectively.}
         \label{histRAVE}
   \end{figure}

    \begin{figure}
   \centering
       \includegraphics[width=0.38\textwidth]{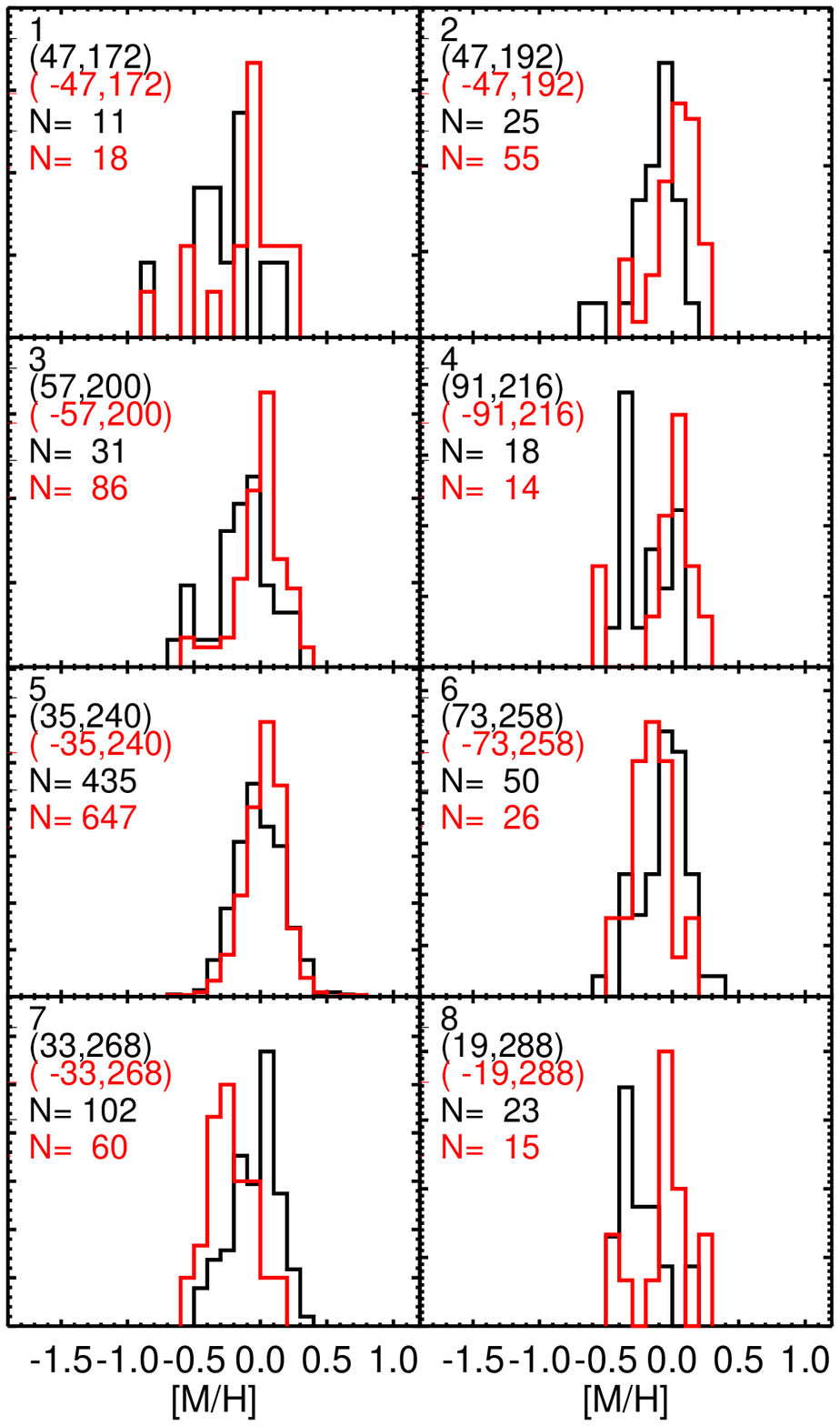}
  \caption{Same as Fig.~\ref{histRAVE} but for GCS. The panels are numbered as in the right bottom panel of Fig.~\ref{metvrvphi}.}
         \label{histGCS}
   \end{figure}

Figures \ref{histRAVE} and \ref{histGCS} show the metallicity distributions for the pairs of symmetric points of Tables~\ref{tRAVE} and \ref{tGCS}, respectively. The difference in the median metallicity between the black and red distributions is evident. We see also differences in the shape of the distributions themselves (e.g., for pairs {3, 5, 6 and 11} in the RAVE sample). 
Note that some histograms might be affected by small number statistics, although they all led to statistically significant differences ($95\%$) in their distributions.


 \section{Metallicity asymmetries as a function of velocity}\label{assymvphi}

       \begin{figure}
   \centering
\includegraphics[width=0.38\textwidth]{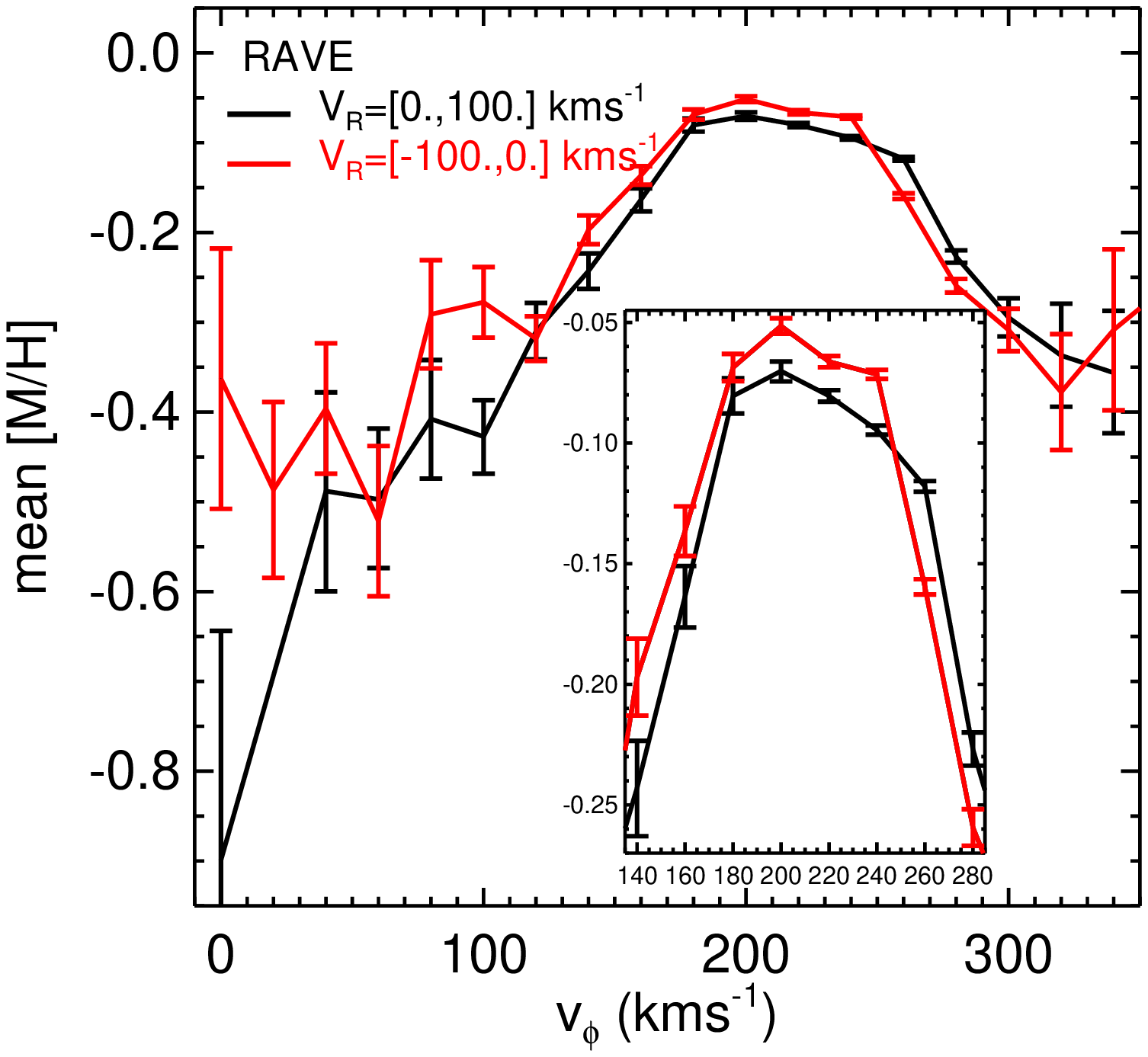}

\includegraphics[width=0.38\textwidth]{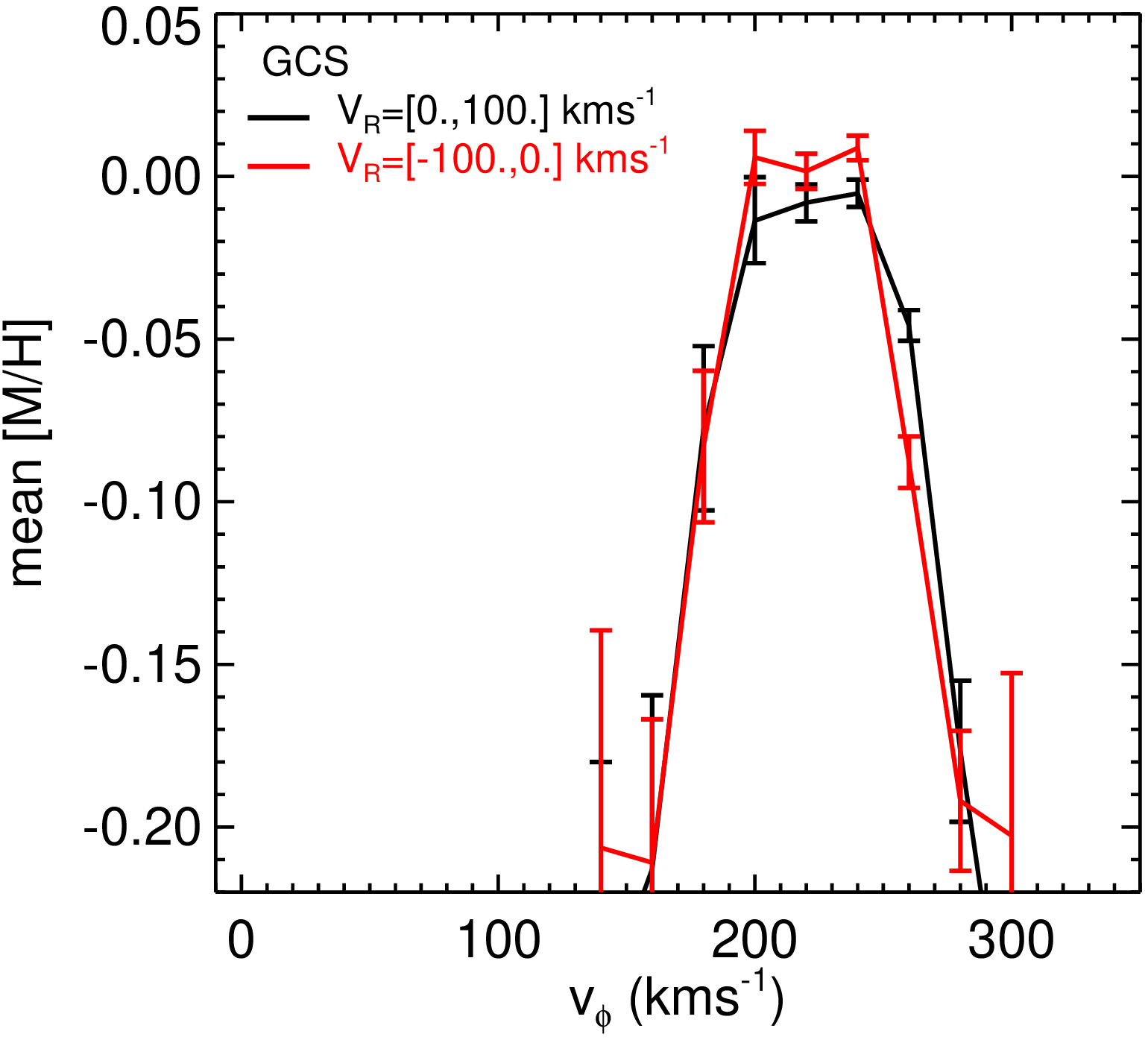}
  \caption{{\em Top:} Mean metallicity as a function of $\vphi$ for the positive and negative $\vr$ (black and red lines, respectively) for the RAVE sample. The error bars show the $75\%$ confidence band. {The box in the top panel shows a zoom around $\vphi=200\kms$.} {\it Bottom:} Same for the GCS sample. }  
  \label{metvphi}
\end{figure} 


 From Fig.~\ref{metvrvphi} and  Tables~\ref{tRAVE} and \ref{tGCS}, it transpires that  $\mh_1-\mh_2$ takes predominantly negative values for low $\vphi$ and positive values for large $\vphi$.
We analyse this trend here. We divide the velocity plane in two parts for positive and negative $\vr$: ($[-100,0]$ and $[0,100]\kms$) and take bins in $\vphi$ of $\Delta\vphi=20\kms$. Figure~\ref{metvphi} compares the mean metallicity as a function of $\vphi$ for {positive $\vr$} (black) and {negative $\vr$} (red) for the RAVE (top) and GCS (bottom) samples. The error bars mark the $75\%$ confidence limits instead of the $95\%$ used in previous sections. {We only plot bins with at least 10 stars.}


{The curves in Fig.~\ref{metvphi} show a clear difference of the mean $\mh$  as a function of $\vphi$ (see Sect. \ref{gradients}), as well as a clear difference between the metallicities for the $\vr<0$ stars and the $\vr>0$ ones. The $\vr<0$ stars are more metal-rich than the $\vr>0$ ones, except for the stars with $\vphi>240\kms$ where the contrary happens.}
For RAVE, the absolute differences {(where they are significant)} are between {$0.01$} and {$0.5\dex$}, with a {median} of {$0.03\dex$}. We see the same trend for the GCS, although it is not so significant. The differences are small but present at $75\%$ confidence. 

We have checked that assuming a different $\Us$ does not affect significantly our results. 
A change in $\Us$ can never smooth out completely the metallicity differences in the whole range of $\vphi$: a $\Us$ {smaller than the one assumed} diminishes the metallicity discrepancies for large $\vphi$ but increases them for low $\vphi$, and the contrary when we assume a larger $\Us$. {Moreover}, to smooth out completely {one of the regimes}, values up to $\Us\sim25\kms$ or down to $\Us<-5\kms$ are necessary. We are not aware of any measurements of $\Us$ of these magnitudes.

\begin{figure}
   \centering
  \includegraphics[width=0.38\textwidth]{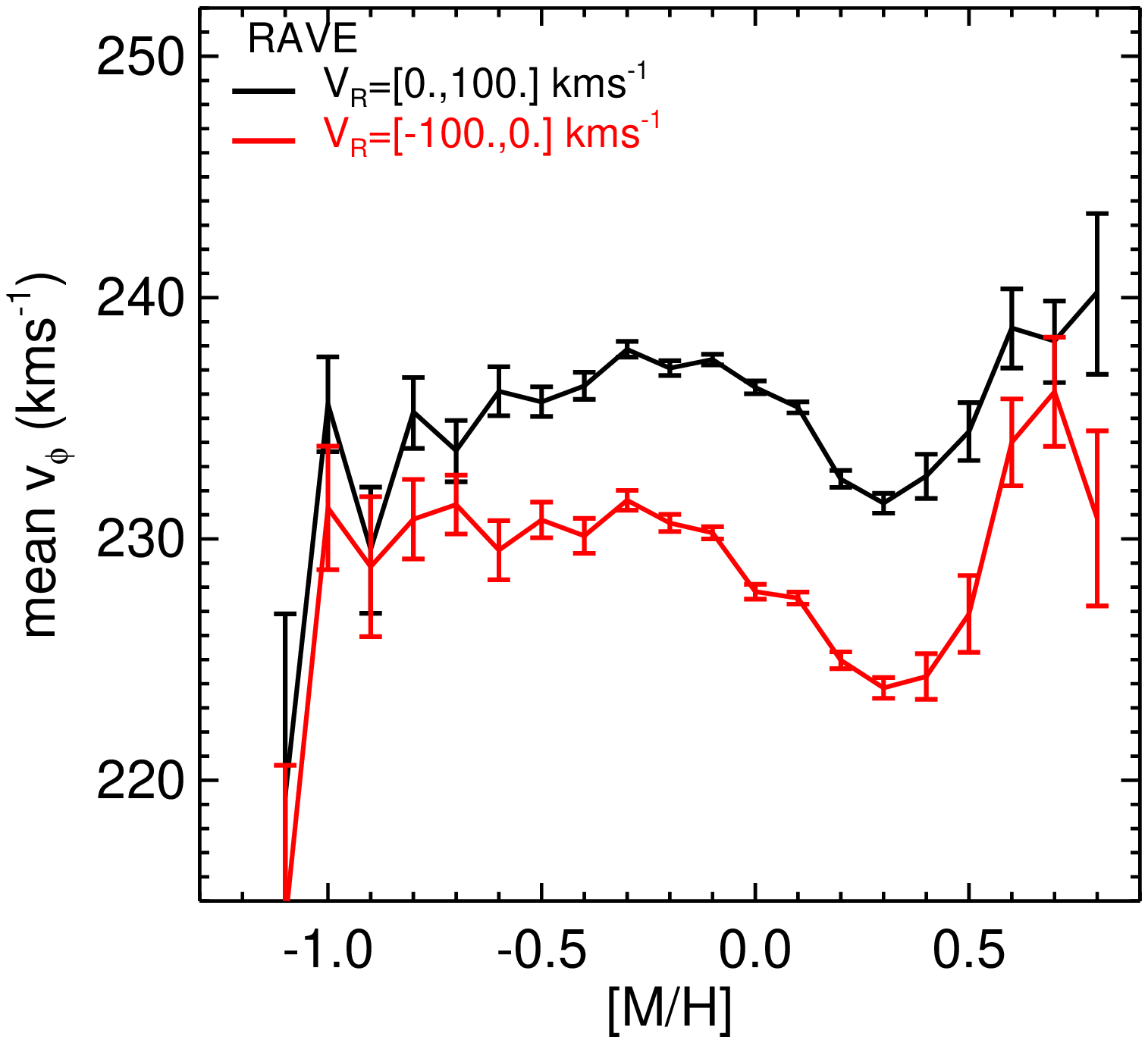}
 \includegraphics[width=0.38\textwidth]{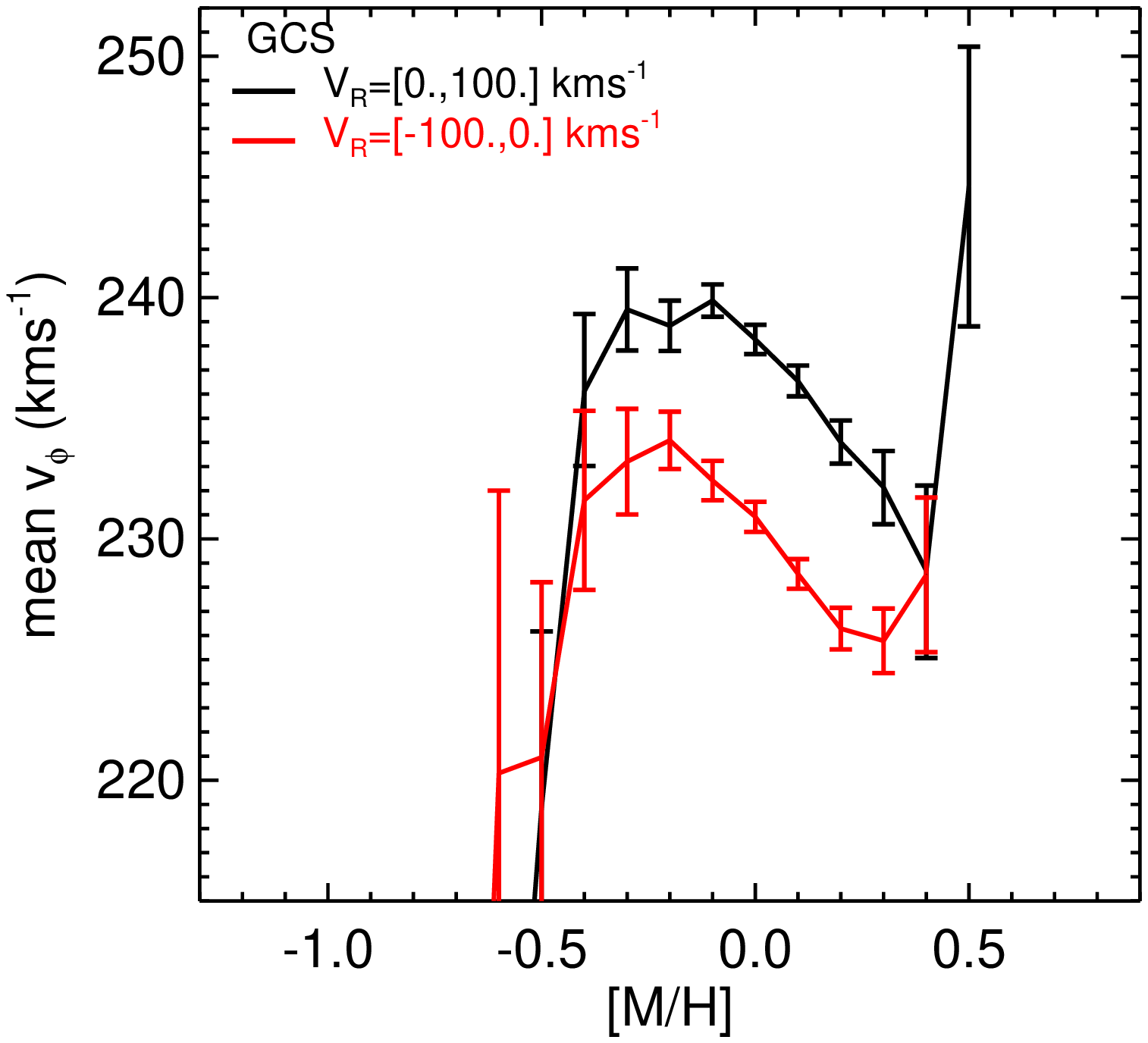}
  \caption{{\em Top:} Mean azimuthal velocity as a function of metallicity for  positive and negative $\vr$ (black and red lines, respectively) for the RAVE sample. The error bars show the $75\%$ confidence band. {\it Bottom:} Same for the GCS sample.}
\label{vphimet}
\end{figure} 

In Fig.~\ref{vphimet} we plot $\vphi$ as a function of $\mh$ for the stars in the {same ranges as before ($\vr=[-100,0]$ in red, $\vr=[0,100]\kms$ in black)} that is the reverse of Fig.~\ref{metvphi}. {We only plot bins with at least 10 stars.} In this plot {stars with $\vr>0$} have a  higher {mean} azimuthal velocity $\vphi$ at fixed metallicity. In this case we only observe one of the regimes that we saw in Fig.~\ref{metvphi}, that is the part for high $\vphi$. This is because the two regimes coexist for a certain metallicity, {but} the part for high $\vphi$ dominates because it has far more stars than the low $\vphi$ part. The difference in velocity for a fixed metallicity ranges from {4.4} to {$9.4\kms$} with a {median} of {$7.2\kms$} for the RAVE sample {and similar values for the GCS}. 
Note how, even though there were only two bins with significant differences in Fig.~\ref{metvphi}, now the trend appears to be very significant in  Fig.~\ref{vphimet}.



%


 \section{Metallicity and azimuthal velocity gradients}\label{gradients}

In Figure~\ref{metvphi}, {independently of } $\vr$ and for both samples, the metallicity increases with $\vphi$ for low $\vphi$ ($<190\kms$), there is a flat part around $\vphi\sim200\kms$, and a decrease for $\vphi>240\kms$. This general behaviour can be linked to the {known} positive and {negative correlations between metallicity and azimuthal velocity for the thick and thin disks, respectively.} 
The left part of Fig.~\ref{metvphi} (low $\vphi$) is dominated by the thick disk, the right part (high $\vphi$) by the thin disk, while the middle part is a mix of both components. 

To measure the gradient between the rotational velocity $\vphi$ and the metallicity, we separate the samples in the two regimes where the slope is positive ($\vphi<190\kms$) and negative ($\vphi>230\kms$). 
Usually the inverted gradient is reported. In that case {for RAVE} we obtain {$43.0\pm 1.0\kms/\dex$}  and {$-8.4\pm 0.2\kms/\dex$}, respectively. These values are similar to the ones in \citet{Wojno2016} who used also RAVE data to separate in a probabilistic way the two disks. The small differences might be due to a different selection of our sample (e.g. they took stars with $SNR>80$ and distances $<1\kpc$), the fact that here we compute the gradient with unbinned data and that they took $\feh$ instead of $\mh$. For the GCS the gradients are higher:  $76 \pm 6\kms/\dex$ and $-18.1\kms/\dex$. 

In Fig.~\ref{vphimet}, we see that the gradient in the left part of the plot (low metallicities) is not very pronounced and noisy for up to $\mh=-0.2$. This might be due to the two regimes (positive and negative gradient) coexisting in this metallicity range. For the range of {$\mh=[$${-0.35,0.25}$$]\dex$} the data show a negative gradient of {$-10.1 \pm  0.5\kms/\dex$}. At higher metallicities ($\mh>$${0.25}$$\dex$), we detect for the first time a positive gradient with slope of {$18 \pm 2 \kms/\dex$} 
(see Sect.~\ref{subsec:vphi_meta} {for an interpretation}). {The lack of stars beyond $0.5\dex$ in the GCS sample impede us from studying this trend {in this sample} but it can be perceived in the last bins.}

 
  \section{Summary, discussion and conclusions}\label{concl}

\subsection{Metallicity asymmetries in the velocity plane}\label{concl1}

{By studying the metallicity as a function of velocity of the solar neighbourhood in RAVE and GCS we have found that:} 
\begin{itemize}
\item The { variation of metallicity over the} velocity plane is highly structured. Most of the main velocity over-densities have a distinct metallicity compared to its velocity surroundings. This is the case of the Hercules and Hyades streams, and of other velocity branch-like structures. 
\item A considerable part of the velocity plane shows significant metallicity differences between $(\vr,\vphi)$ and its symmetric region $(-\vr,\vphi)$. {We obtain similar results with the two independent samples, which confirm the robustness of our findings.}
The typical {metallicity} differences are of 
{$0.05\dex$} and $0.12\dex$, for RAVE and GCS, respectively, with values up to $0.6\dex$ {and at $95\%$ confidence}. 

\item For low azimuthal velocity $\vphi$, stars with negative $\vr$ (i.e., stars moving outwards in the Galaxy) {have on average higher metallicity than those moving inwards}. This region coincides with the Hercules and Hyades streams. On the contrary, {for stars with higher $\vphi$, the ones moving inwards (positive $\vr$) have higher metallicity than for negative $\vr$}. The limit between the two regimes is $\vphi\sim250\kms$. 
\end{itemize}


The asymmetric metallicity distribution in the velocity plane that we have found {implies that there are sub-populations that are not yet well phase-mixed and/or that there is an effect of the non-axixymmetric parts of the potential.}

{For structures like Hercules or Hyades, we rule out the first hypothesis since }it has been already shown that the
metallicity, age and even mass distributions of their stars are not compatible with being remnants of a
 sub-population \citep{Raboud1998,
  Bensby2007, Bensby2014, DeSilva2011, Pompeia2011, Famaey2007}.
We find a few small regions of the velocity distribution
that present a distinct metallicity and do not correspond to any
known velocity group, and could potentially be remnants of { a} cluster or a disrupted
satellite. A detailed chemical analysis is required to confirm
this.

%

Our favoured interpretation of the global asymmetries in metallicity is that they are due to the {non-axisymmetries of the potential (e.g. bar and spiral arms). In this case, stars following orbits with the same $\vphi$ but opposite signs of $\vr$ would not have necessarily the same guiding radii, and thus, could have different metallicities given by the metallicity gradient of the disk} {(see references in Section~\ref{intro})}.

  \begin{figure*}
   \centering
\includegraphics[height=0.3\textheight]{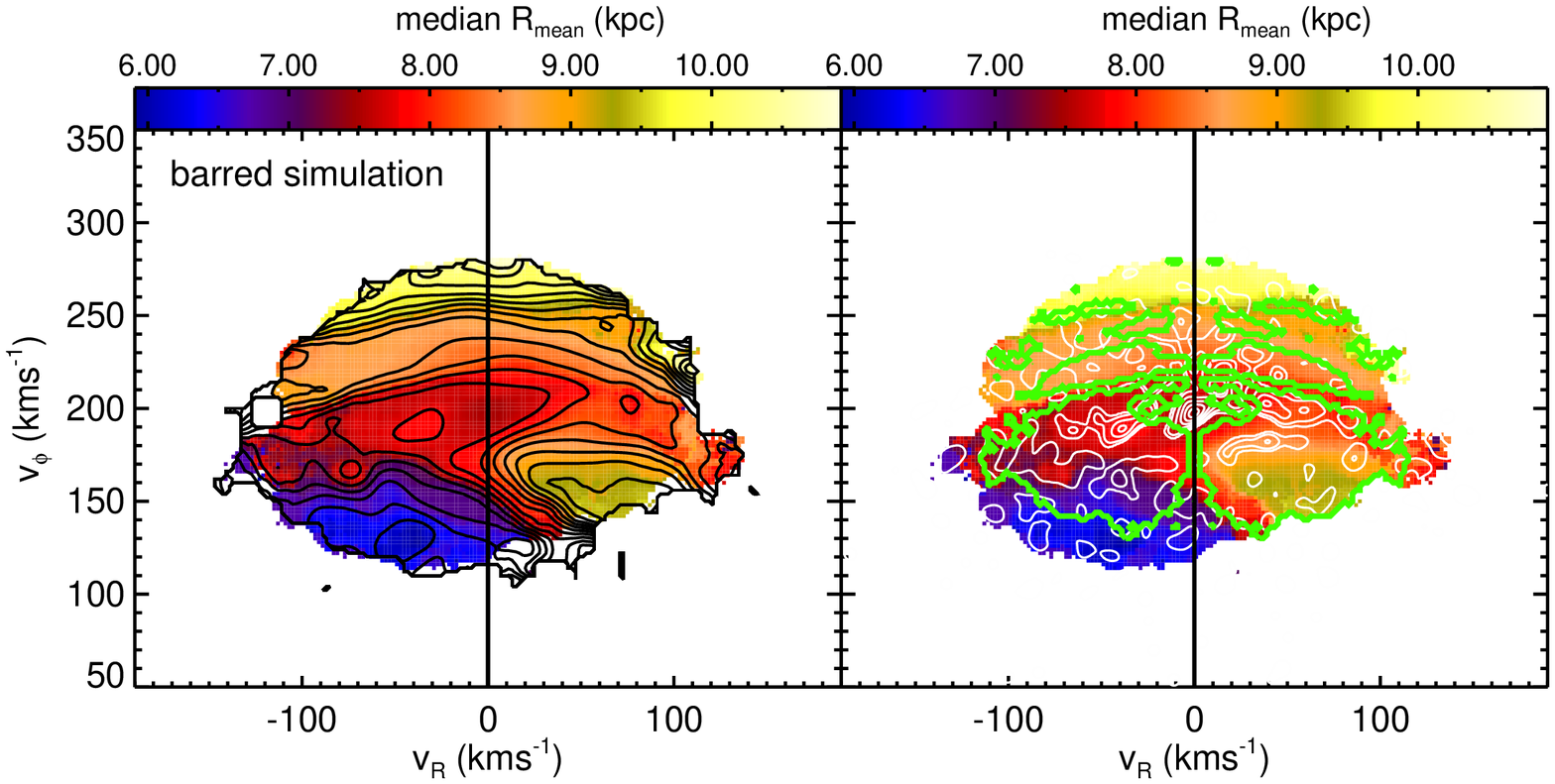} \hspace{-1.6cm} 
\includegraphics[height=0.3\textheight]{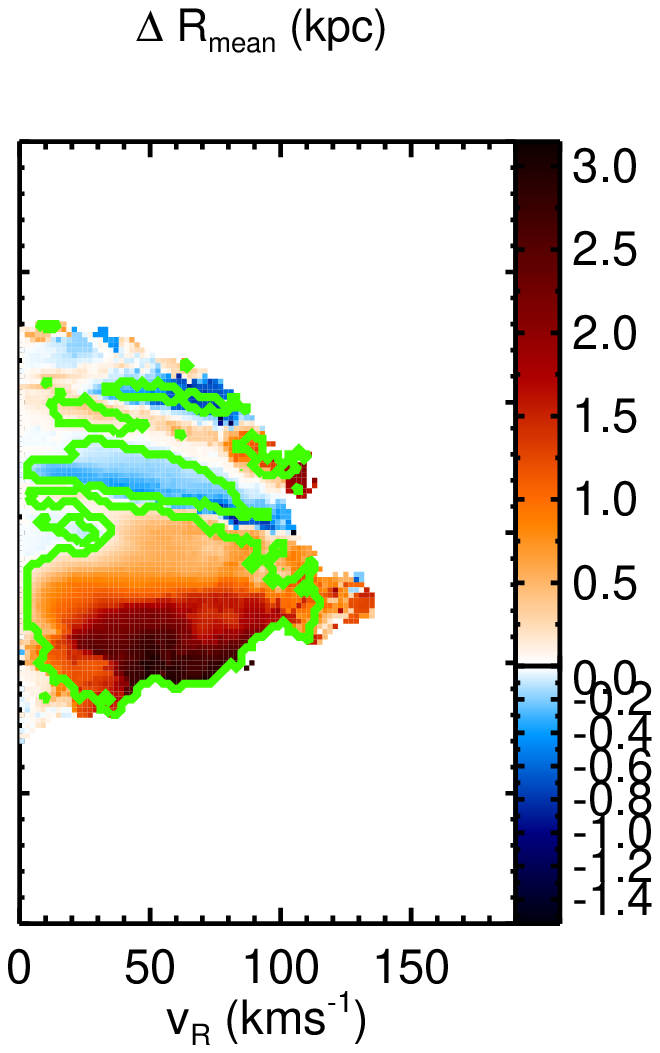}

\caption{Median orbital radius of the bins in the velocity plane of simulations in a Galactic potential with a bar. 
The {\it left} { panel shows contours every $0.25\kpc$ in the range of $[6,10.5]\kpc$.} { The {\it middle} panel is the same as the left one but shows} the contours of the wavelet transform (white contours), which indicate the main over-densities in the velocity plane, and the green contours that mark the regions where the median orbital radius is significantly different with a $95\%$ confidence from its symmetric $\vr$ region. The {\it right} panel shows the differences in the median orbital radius (depicted in the left and middle panels) for particles with positive and negative $\vr$. The green contours {are the same as for the middle panel.} 
}        
  \label{Rvrvphi}
\end{figure*} 

{We use a} simulation from \citet{Monari2015,Monari2016b} to support this interpretation. {The simulation consists of} orbital integrations with an analytic potential for the Milky Way containing a central bar\footnote{The bar is modelled as a quadrupole (e.g., \citealt{Dehnen2000}) with a pattern speed of $\Omega_\mathrm{p}=1.89\Omega(\Rsun)$, consistent with the estimate of \citet{Antoja2014}. The amplitude of the bar potential is null at the beginning of the simulation, grows for 3 Gyr and is kept constant for the following 3\,Gyr. The local volume is placed at an angle of $-25\deg$ with the respect of the bar's long axis, in the direction opposite to the bar's rotation.}.
 Figure~\ref{Rvrvphi} (left) shows the kinematics of stars in a solar neighbourhood-like volume {colour-coded by mean orbital radius over the time when the bar is fully grown}.  
This plot is equivalent to Fig.~\ref{metvrvphi} {with mean orbital radius} instead of metallicity. The scale of the colour bars is such that smaller guiding radii have blue colours like higher metallicities (as would correspond to a negative radial metallicity gradient in the disk). In Fig.~\ref{Rvrvphi} we see a dependence on $\vr$ that makes the distribution of mean radius asymmetric with respect to the $\vr=0$ axis\footnote{In an axisymmetric model, we would see a colour pattern that changes with $\vphi$: orbits with small $\vphi$ are in their apocentres and have, thus, a smaller orbital radius, while orbits with high $\vphi$ are in their pericentres which corresponds to larger mean orbital radius.}. The differences are statistically significant: the green contours {in the middle panel} show the regions where the radius is different with a $95\%$ confidence to its symmetric point in the velocity plane. This can be better seen in the right panel, where we subtract the left part of the velocity plane from the right part as we did in Fig.~\ref{metvrvphi}. Indeed, we obtain a very similar colour pattern compared to Fig.~\ref{metvrvphi}. For low $\vphi$, the {stars currently} moving inwards ($\vr>0$) come from  outer radii ($\Delta R_{mean}>0$, red colours) and would have a lower metallicity {compared to stars with $\vr<0$}. For high $\vphi$, the orbits moving inwards ($\vr>0$) come from  inner radii ($\Delta R_{mean}<0$, blue colours) and would have to a higher metallicity. This model is simple and {does not include} the ages of stars, but it gives a basic explanation of our findings relating them to the orbits induced by the bar of the Galaxy.

The median difference in orbital radius where there are significant differences in Fig.~\ref{Rvrvphi}  is $0.7\kpc$, with a range of $[0.01,3.1]\kpc$. 
Given a metallicity gradient of $-0.07\dex/\kpc$, {this  corresponds to a} metallicity difference of $0.05\dex$. This is similar to the median differences in RAVE, but smaller than for GCS. Note that this is only an order of magnitude estimate: the metallicity gradients in our Galaxy are very uncertain and even more their values in the past
 \citep[but see][and the models by \citealt{Minchev2013}]{Yong2012}.
The simulation does not reproduce the small-scale metallicity
structure of Fig.~\ref{metvrvphi} but only the global asymmetry as
function $\vphi$. The granularity in the data could be explained by
the presence of tracers of clusters, satellite remnants or other
orbital effects\footnote{In our model the asymmetric distribution of stars in $v_R$ is caused by the bar. However, also the  spiral arms can induce strong kinematic imprints, depending on the {location, pattern speed and mass} of the arms  \citep[e.g.][]{Antoja2011}. We have also assumed the {bar's} outer Lindblad resonance is in proximity of the
  Sun, as in the classical estimates of its angular velocity
  \citep[e.g.][]{Binney1997,Bissantz2003}. These estimates
  are, however, challenged by recent models of gas dynamics and star
  counts in the inner Galaxy, which imply a much slower pattern speed
   \citep{Wegg2015,Portail2015,Li2016}.}.
  
Our results { caution} { against} computing orbital eccentricities,
guiding radii and other {stellar} orbital quantities using axisymmetric
potentials. We see that in our simulation the orbits are not symmetric
in $\vr$. We observe differences of up to $3\kpc$ in the mean orbital
radius, which is not negligible.

The {metallicity asymmetry that}  we find in the RAVE and GCS samples is very similar to the asymmetries in the velocity plane found in \citet{Antoja2015}.  { In that study we saw an unbalanced number of stars for positive and negative $\vr$ and also} differences in the mean $\vr$ {as a function of $\vphi$}. The transition point was at a very similar $\vphi$ {compared to the present work and also the effects of the bar were suggested as a possible explanation}. {The global scenario would be the following.}  The bar creates an over-density of stars at $\vr<0$ ({currently} moving outwards) for low $\vphi$ that explains the existence of the Hercules stream \citep[e.g.][]{Dehnen2000,Antoja2009,Antoja2014}. This stream {is composed by} stars following orbits that come from the inner Galaxy (thus, more metallic) compared to the stars at the same $\vphi$ but positive $\vr$. Additionally, {the bar forms an} over-density of stars at $\vr>0$ ({currently} moving inwards) for high $\vphi$ that {is made of stars that} on average come from inner regions of the Galaxy (thus, more metallic) compared to the stars at the same $\vphi$ but negative $\vr$.

Some studies reported
azimuthal metallicity gradients in the stellar component, e.g., in
clusters \citep{Davies2009}  and Cepheids \citep{Lepine2011, Luck2011}. These
were associated to a patchy star formation driven by the
 bar and the incomplete mixing of metallicity in azimuth after star formation on the spiral arms,
respectively.  Simulations such as  in
 \citet{Dimatteo2013,Grand2016, Miranda2016} also predict stellar azimuthal
metallicity gradients due to radial migration.  The connection
between these gradients and {the metallicity asymmetry that we find} needs to be
investigated.

\subsection{Comparison with previous studies}

Most of the previous studies found that the majority of the over-densities show large metallicity dispersion and distributions compatible with the general field population (see Sections~\ref{intro} and \ref{concl1}). 
{Because of this,} 
\citet{Bovy2010} concluded that these over-densities are more likely associated with transient perturbations. They found that the Hyades moving group is the only one with evidence for a higher {metallicity than the one corresponding to its mean orbital radius in an axisymmetric model. This is consistent with an origin related to the inner Lindblad resonance of the spiral arms as suggested in  \citet{Quillen2005}.} This {agrees} with our results where the Hyades group is more metal rich than its symmetric counterpart. 
Regarding the Hercules stream, \citet{Bovy2010} and \citet{Ramya2016}  found that it shows only weak evidence of a higher metallicity than average, which would not be in agreement with being due to the bar's outer Lindblad resonance.  \citet{Liu2016} identified a structure in the age-metallicity plane, that they called {\it narrow stripe}, in the LAMOST K giant sample, which is composed by the Hercules stream and other groups. They found that this structure is made of stars with a small guiding radius ( $\sim6\kpc$) but that it has an age-metallicity distribution that is consistent with stars being formed at a radius of $4\kpc$. This made them conclude that these stars might have migrated from the very inner parts of the Galaxy instead of followed the resonances of the bar. {However, we find that the orbits of a barred potential   with median radii of $\sim 6-7.5\kpc$ are sufficient to explain the metallicity differences of Hercules and its $\vr>0$ symmetric region, without any need of additional radial migration.}

\subsection{The azimuthal velocity-metallicity gradients}
 \label{subsec:vphi_meta}
We observe a positive azimuthal velocity-metallicity gradient for low $\vphi$ and a negative one for high $\vphi$. This is in agreement with the positive and negative gradients found for thick and thin disk, respectively  
\citep{Kordopatis2011b, Kordopatis2013b, Kordopatis2013c, Kordopatis2015b, Lee2011, Liu2012b, Spagna2010, Haywood2013,   RecioBlanco2014, Wojno2016}.
The slopes that we measure are {similar} to previously reported values. The small differences could be due to the different weight of the population in each of the two regimes. 


{ In the RAVE sample} we find a new regime in the velocity-metallicity gradient: a positive gradient for high metallicity, connecting with the negative gradient of the thin disk.  This is only evident when putting $\mh$ on the x-axis, as the metal-rich tail only represents a small fraction of all stars at a given azimuthal velocity. Although it was not mentioned there, we see signs of the same gradient in the simulation of \citet{Loebman2011} for intermediate ages and their analysis of the GCS. This {positive gradient} could be a signature of radial migration mechanisms, where 
 these metal-rich stars are migrators from the inner disk, with  smaller velocity dispersions (i.e. small eccentricities) than the global population. This would naturally lead to a smaller asymmetric drift,  putting them closer to the circular velocity of the Local Standard of Rest. This is consistent with the small eccentricities found for the super metal-rich stars in  \citet[][see their Figs.~6 and 9]{Kordopatis2015}.

\vspace{0.3cm}
We have shown that the chemical measurements {show asymmetries in the velocity plane which suggest the presence of incomplete phase-mixing and/or the effects of spiral arms and the bar on stellar orbits.}
Very soon with  Gaia  \citep{Perryman2001,deBruijne2012} and its follow up surveys (WEAVE,  4MOST) we expect to disentangle the exact origin of each velocity over-density, to solve current model degeneracies, and to extend the study to different parts of the disk, thus building a global chemo-dynamical model of the Milky Way disk.


  


\begin{acknowledgements}
 {We thank the referee, Professor Binney, for the careful reading and advice.}
 TA is supported by an ESA Research Fellowship
in Space Science. {This work was supported by the MINECO (Spanish Ministry of Economy) - FEDER through grant ESP2016-80079-C2-1-R and ESP2014-55996-C2-1-R and MDM-2014-0369 of ICCUB (Unidad de Excelencia 'Maria de Maeztu').} GM is supported by a postdoctoral grant from the {\it Centre National d'Etudes Spatiales} (CNES). AH acknowledges financial support from a VICI personal grant from the Netherlands Organisation for Scientific Research.   We thank the CEA and Nice Observatory for the MR software. Funding for RAVE has been provided by: the Australian Astronomical Observatory; the Leibniz-Institut fuer Astrophysik Potsdam (AIP); the Australian National University; the Australian Research Council; the French National Research Agency; the German Research Foundation (SPP 1177 and SFB 881); the European Research Council (ERC-StG 240271 Galactica); the Istituto Nazionale di Astrofisica at Padova; The Johns Hopkins University; the National Science Foundation of the USA (AST-0908326); the W. M. Keck foundation; the Macquarie University; the Netherlands Research School for Astronomy; the Natural Sciences and Engineering Research Council of Canada; the Slovenian Research Agency; the Swiss National Science Foundation; the Science \& Technology Facilities Council of the UK; Opticon; Strasbourg Observatory; and the Universities of Groningen, Heidelberg and Sydney.
The RAVE web site is at https://www.rave-survey.org.

\end{acknowledgements}

\bibliographystyle{aa} 
\bibliography{mybib}

\end{document}